\documentclass[sigconf]{acmart}

\usepackage{subcaption}
\usepackage{url}
\usepackage{xcolor}
\usepackage{multirow}
\usepackage{graphicx}
\usepackage{fontawesome5}

\title[Empirical and Sustainability Aspects of SE Research in the Era of LLMs: A Reflection]{Empirical and Sustainability Aspects of Software Engineering Research in the Era of Large Language Models: A Reflection}

\author{David Williams}
\affiliation{
  \institution{University College London}
  \city{London}
  \country{United Kingdom}}
\email{david.williams.22@ucl.ac.uk}

\author{Max Hort}
\affiliation{
  \institution{Simula Research Laboratory}
  \city{Oslo}
  \country{Norway}}
\email{maxh@simula.no}

\author{Maria Kechagia}
\affiliation{
  \institution{University of Athens}
  \city{Athens}
  \country{Greece}}
\email{makechag@ba.uoa.gr}

\author{Aldeida Aleti}
\affiliation{
  \institution{Monash University}
  \city{Melbourne}
  \country{Australia}}
\email{aldeida.aleti@monash.edu}

\author{Justyna Petke}
\affiliation{
  \institution{University College London} 
  \city{London}
  \country{United Kingdom}}
\email{j.petke@ucl.ac.uk}

\author{Federica Sarro}
\affiliation{
  \institution{University College London}  
  \city{London}
  \country{United Kingdom}}
\email{f.sarro@ucl.ac.uk}

\renewcommand{\shortauthors}{Williams, et al.}

\date{April 2026}

\renewcommand\footnotetextcopyrightpermission[1]{}

\acmConference[ICSE-NIER '26]{2026 IEEE/ACM 48th International Conference on Software Engineering}{April 12--18, 2026}{Rio de Janeiro, Brazil}


\begin{document}

\begin{abstract}
Software Engineering (SE) research involving the use of Large Language Models (LLMs) has introduced several new challenges related to rigour in benchmarking, contamination, replicability, and sustainability. In this paper, we invite the research community to reflect on how these challenges are addressed in SE. Our results provide a structured overview of current LLM-based SE research at ICSE, highlighting both encouraging practices and persistent shortcomings. We conclude with recommendations to strengthen benchmarking rigour, improve replicability, and address the financial and environmental costs of LLM-based SE.
\end{abstract}

\maketitle

\section{Introduction} \label{sec:intro}
Despite the potential of Large Language Models (LLMs) to augment human capabilities and streamline software development workflows, their integration into Software Engineering (SE) research raises several critical concerns. For instance, benchmarking should be performed systematically to ensure fair evaluation, particularly given how potential data leakage from training corpora can skew performance metrics~\cite{sainz-2023-nlp}. Reproducibility and replicability practices should become mandatory to address the stochastic nature of LLMs and the lack of standardised evaluation protocols~\cite{sallou2024breaking}. Further, efforts should be made to reduce the computational costs of training and inference, both in terms of energy and infrastructure, which pose significant barriers to accessibility and sustainability~\cite{Hort2023}. Addressing these challenges is crucial to harnessing the full potential of LLMs while maintaining scientific rigour, ethical standards, and equitable access.

To understand how our community is tackling the aforementioned challenges, we analyse the empirical research literature published in the research track of the top SE conference, the International Conference on Software Engineering (ICSE), between 2023 and 2025. We identified 177 relevant papers and manually extracted several key pieces of information, including types of models used, experimental protocols, benchmarks, reported costs, and other details relevant to empirical studies. We manually analysed this information to establish:

\textbf{RQ1. Models \& Benchmarking.} \textit{Which LLMs are used in SE research and how are they benchmarked?} We aim to assess whether current studies provide reliable evidence of improvement over non-LLM techniques and to highlight gaps where more rigorous evaluation protocols are needed. We check (i) whether open or closed models are used, (ii) model families, (iii) comparisons with non-LLM baselines, and (iv) targeted programming languages.

\textbf{RQ2. Contamination.} \textit{How well do authors tackle the problem of data leakage/contamination?} Since data leakage and contamination directly threaten the validity of empirical findings, we investigate (i) whether authors mention data contamination as a threat and (ii) any strategies proposed to mitigate contamination.

\textbf{RQ3. Replicability. }\textit{How replicable are LLM-based studies?} Replicability underpins scientific credibility. Given the stochastic nature of model outputs, opaque APIs, and frequent model updates, we identify (i) whether LLM configuration details are reported and (ii) whether artefacts are shared and recognised with a badge.

\textbf{RQ4. Sustainability.} \textit{What are the costs of LLM-based SE research?} Training and deploying LLMs is resource-intensive, with high financial and environmental costs. By examining how sustainability is addressed, we aim to raise awareness of these hidden costs and encourage the community to adopt more transparent, responsible practices that balance innovation, accessibility, and environmental responsibility. We check (i) which experimental costs are reported, (ii) the costs of LLM-based SE research, and (iii) how they are sustained. We support these findings with a discussion of 57 responses to a questionnaire about experiment costs, which we distributed to authors of the 177 relevant papers. 

\section{Methodology} \label{sec:met}

To answer our RQs, we used a mixed-methods approach, integrating quantitative and qualitative data gathered through a literature review and an author questionnaire. We incorporated temporal analyses, where possible, to analyse how trends have shifted.

\textbf{\textit{Article Search.}}
The study was conducted on the 692 articles published in the ICSE technical research track in 2023, 2024, and 2025, a period coinciding with the widespread adoption of LLMs after the release of ChatGPT in late 2022. As the premier venue in SE, ICSE sets the bar for methodological rigour and research quality in the field, meaning our insights can be regarded as a reliable reflection of the community’s current priorities and practices.

First, we filtered the initial 692 papers based on the following keywords: ``\textit{LLM}'', ``\textit{language model}'', ``\textit{conversational AI}'', ``\textit{chatbot}'', ``\textit{genAI}'', ``\textit{generative}'', ``\textit{AI-assisted}'', ``\textit{foundation model}'', ``\textit{BERT}'', ``\textit{encoder}'', ``\textit{decoder}'', ``\textit{autoencoder}'', ``\textit{transformer}'', and ``\textit{agent}'', retrieving 304 candidates. Among these, we manually selected relevant papers matching the following inclusion criteria: \textit{articles that present an empirical experiment involving at least one LLM, either as the main approach or as a baseline}. For instance, we exclude papers that focus solely on qualitative insights from LLM user studies. When referring to LLMs, we use the definitions proposed by Fan et al.~\cite{fan-2023-survey}, namely encoder-only (e.g., BERT, DeBERTa, RoBERTa), encoder-decoder (e.g., T5, BART), and decoder-only (e.g., GPT, LLaMA, Claude). Our selection followed a three-step process widely adopted in previous surveys (e.g. ~\cite{martin2016survey,hort21tse,10.1145/3533767.3534405}): (i) We excluded all papers whose titles clearly did not meet our inclusion criteria; (ii) next, we checked abstracts; (iii) finally, we reviewed the full content of articles remaining from the previous two steps and excluded those that did not meet our inclusion criteria. This process resulted in 174 papers selected for data extraction. To mitigate the risks of missing relevant papers, we (i) checked whether we missed any keywords by computing word clouds from the abstracts of the 174 papers (see ~\cite{repo} for full results), finding no additional relevant keywords; (ii) randomly sampled 20\% of the 388 papers not selected by our keyword search for each year. This sample comprised 80 papers (32 for 2023, 28 for 2024, and 20 for 2025), which we manually inspected using the process above. We found only three false negatives and included them in the survey. Thus, we analyse 177 papers in total, with the complete list available in our online repository~\cite{repo}.

\textbf{\textit{Data Extraction.}}
\label{sec:data_extraction}
Next, we performed data extraction guided by the RQs. We minimised the threat of data extraction and interpretation bias by standardising the process for extracting and consolidating data from the papers. We minimised the threat of reviewer bias by involving at least two authors at each step of this study. For each RQ, we identified the specific information to be collected from the selected papers. To operationalise this, each author was assigned a subset of papers and independently extracted the required information. For reliability, we began with a pilot phase where each author analysed four papers, then met to discuss the extraction process, clarify ambiguities, and refine the extraction schema. This process was repeated for a new set of four papers, ensuring ambiguities were resolved before analysing the full set.

Once the extraction from all 177 papers was completed, the information was consolidated into a unified taxonomy. This consolidation was performed iteratively: one of the authors compared entries across papers, standardised terminology, and grouped similar items under common categories. Any uncertainties or disagreements were resolved collaboratively among all authors to ensure consensus. Finally, the resulting taxonomy provided a structured view of how the ICSE community addresses benchmarking, contamination, replicability, and sustainability, which served as the basis for our subsequent analysis.

\textbf{\textit{User Study.}}
To better understand the costs incurred by the SE research community for using LLMs in empirical studies, we developed a brief questionnaire and distributed it to the authors of relevant papers included in our survey. This consisted of 10 questions aimed at gaining insights into the costs incurred by researchers using LLMs over the past year, how these costs were covered, and their expectations regarding both aspects for the next 12 months. Due to space constraints, the full questionnaire is available online~\cite{repo}. This study received approval from the ethics committee of the Computer Science Department at UCL (Project ID: 1819).

\section{Results \& Reflections} \label{sec:res}
We analysed 177 papers reporting empirical SE studies using LLMs: 32 from ICSE 2023, 55 from ICSE 2024, and 90 from ICSE 2025. Although it is expected that the number of such studies would have grown over the years, we observe that these studies accounted for 2.4 times as many articles published in the ICSE research track in 2025 compared to 2023 (36.6\% vs. 15.2\%).

\noindent \textbf{RQ1. Models and Benchmarking}
\label{sec:res_models}
\hfill \break
\indent \textbf{\textit{Open vs. commercial models.}} Our analysis reveals that 136 out of the 177 relevant papers (76.8\%) use open models (e.g., DeepSeek, CodeT5), whereas 106 out of 177 papers (59.9\%) consider closed models (e.g., OpenAI's GPT models, Google's Gemini, Anthropic's Claude). Table~\ref{tab:open_closed} describes the prevalence of open and commercial LLMs used in empirical LLM-based studies. 

\begin{table}
\caption{Num. papers employing open \& commercial models.}
\label{tab:open_closed}
\centering
\begin{tabular}{|l|r|r|r|}
\hline
\textbf{Year} & \textbf{Only Open} & \textbf{Only Comm.} & \textbf{Open \& Comm.} \\
\hline
2023            & 22 / 32 (68.8\%)  & 4 / 32 (12.5\%)   & 6 / 32 (18.8\%)   \\ \hline
2024            & 25 / 55 (45.5\%)  & 13 / 55 (23.6\%)  & 17 / 55 (30.9\%)  \\ \hline
2025            & 24 / 90 (26.7\%)  & 24 / 90 (26.7\%)  & 42 / 90 (46.7\%)  \\ \hline
\textbf{Total}  & 71 / 177 (40.1\%) & 41 / 177 (23.2\%) & 65 / 177 (36.7\%) \\ \hline
\end{tabular}
\end{table}

\faHandPointRight\ Overall, there is an increasing use of closed models over time, particularly in the number of studies that feature only closed or commercial models. We raise this as a concern because commercial models are typically less financially accessible to researchers, and replicability is hindered as these models are frequently deprecated when newer versions are released.

\textit{\textbf{Model Family \& Benchmarking.}}
We examined the number of papers that consider a particular model family. The top 5 families of models used across all three years are: \textbf{1.} GPT-4 (47 papers), \textbf{2.} GPT-3.5 (44 papers), \textbf{3.} CodeBERT (34 papers), \textbf{4.} CodeLlama (26 papers), and \textbf{5.} CodeT5 (22 papers). A full table of the most popular models per year (and additional analysis on how many models were used in various papers) can be found in our online appendix~\cite{repo}. Most notably, we observe a significant shift toward OpenAI’s commercial models from 2023 to 2025: while CodeBERT was the most popular model in 2023 (featuring in 11 out of 32 papers, or 34\%), by 2025 it appears in only 12 out of 90 papers (13\%), ranking fifth, whereas GPT-3.5 and GPT-4 feature in 39 (43\%) and 45 (50\%) papers, respectively.

For each study utilising an LLM to solve a given task, we checked whether the LLM is benchmarked with at least one non-LLM approach. While the number of studies benchmarking their proposal against non-LLM techniques has increased over time, the proportion of studies that do so has decreased. Specifically, 27 out of 32 papers (84.4\%) did so in 2023, 36 out of 55 (65.5\%) in 2024, and 51 out of 90 (56.7\%) in 2025.

\faHandPointRight\ We observe a shift in the technologies adopted, from BERT-based models in 2023 to GPT-based models in 2025. While the adoption of more complex and costly technologies increases, there does not seem to be a corresponding increase in benchmarking them against non-LLM counterparts. This should be prioritised to ensure the adoption of these technologies remains cost-effective.

\textbf{\textit{Targeted Programming Languages.}}
We examined the prevalence of programming languages (PLs) used in LLM-based empirical studies and found that Java was the most popular language overall, followed by Python, C, C++, and JavaScript. Interestingly, Table~\ref{tab:programming_languages} shows that Java is becoming less common in studies over the years (from 43.8\% of studies in 2023 to 37.9\% in 2025), while Python has grown more popular (28.1\% to 31.6\%). In terms of evaluating across multiple PLs, we see more concerning trends. Overall, only 27.1\% of papers evaluate with more than one PL. Notably, only 22.2\% of 2025 papers evaluate with more than one PL, compared to the more positive results of 2023 (31.2\%) and 2024 (32.7\%). This trend is problematic for replicability, as LLMs have been shown to perform better on some PLs than others~\cite{twist2025pref}, implying that results reported on one PL will not necessarily generalise to other PLs or contexts.

\faHandPointRight\ To mitigate the risk of reporting potentially inflated performance estimates, we urge researchers to, where possible, evaluate novel LLM SE techniques on multiple programming languages.

\begin{table}
\caption{Prevalence of the top 5 most popular targeted programming languages for LLM-based empirical studies.}
\label{tab:programming_languages}
\centering
\begin{tabular}{|l|r|r|r|r|}
\hline
\textbf{Language}   & \textbf{2023 - 32} & \textbf{2024 - 55}   & \textbf{2025 - 90} & \textbf{Total - 177} \\
\hline
Java                & 14 (43.8\%)       & 22 (40.0\%)           & 31 (34.4\%)       & 67 (37.9\%)   \\ \hline
Python              & 9 (28.1\%)        & 18 (32.7\%)           & 29 (32.2\%)       & 56 (31.6\%)   \\ \hline
C                   & 7 (21.9\%)        & 11 (20.0\%)           & 12 (13.3\%)       & 30 (16.9\%)   \\ \hline
C++                 & 2 (6.3\%)         & 9 (16.3\%)            & 12 (13.3\%)       & 23 (13.0\%)   \\ \hline
JavaScript          & 4 (12.5\%)        & 3 (5.5\%)             & 6 (6.7\%)       & 13 (7.3\%)   \\ \hline
\end{tabular}
\end{table}

\noindent \textbf{RQ2: Contamination}\hfill \break
\label{sec:res_contamination}
\indent Data contamination or leakage refers to the scenario where a model is evaluated using data present in its training set, which can lead to misleading results and inflated performance estimates~\cite{Golchin25, sainz-2023-nlp}. As LLMs are trained on vast, often closed-source corpora, it can be challenging to determine whether an evaluation dataset is contaminated. Regardless, papers should mention such cases as threats to validity and discuss any strategies considered to address and mitigate these issues. 

We found that only three articles (two published in 2024~\cite{yang2024memorization,alkaswan-2024-memorisation} and one in 2025~\cite{nie2025memorisation}) specifically tackle the problem of contamination/memorisation. Moreover, only 58 out of 177 papers (32.8\%) discuss data contamination as a threat to validity. We observed an upward trend in the number of papers reporting data contamination over the examined period from 2023 to 2025. Specifically, for 2023, contamination is reported in 6 out of 32 papers (18.8\%), 14 out of 55 (25.5\%) for 2024, and 38 out of 90 (42.2\%) for 2025. However, despite the rising trend, the percentage of papers that mention such threats remains below 50\%.

Some papers acknowledged the issue of data contamination without providing strategies to minimise the risk to the validity of the results~\cite{ahmed2024automatic}. In other papers, a common approach to addressing contamination is temporal filtering, which involves collecting data created after the training cutoff date of the evaluated models~\cite{jiang2025rocode,wei2024demystifying,zhang2025instruct,wu2025empirical}. For example, papers use datasets like GitBug-Java~\cite{bouzenia2025repairagent}, JLeaks~\cite{wang2025boosting}, DebugBench~\cite{xu2025aligning}, and ClassEval~\cite{zhang2025instruct} that were constructed specifically to minimise overlap. Another strategy is benchmark design through obfuscation or perturbation, such as OBFUSEVAL~\cite{zhang2025unseen} or DS-1000~\cite{ouyang2025knowledge}, which alters code or phrasing while preserving functionality so that models cannot rely on memorised examples. Data cleaning~\cite{nie2025memorisation}, data splitting~\cite{ding2024vulnerability}, and deduplication are widely used, with clone detection~\cite{yang2024memorization} employed to eliminate duplicates across the training, validation, and test sets. In many cases, experiments are replicated across multiple datasets or with synthetic benchmarks specifically designed to minimise leakage risk. Data contamination is further mitigated by cross-dataset evaluation~\cite{batole2025llm} and ablation studies~\cite{ke2025niodebugger,wang2025boosting,guo2025intention}. By evaluating on multiple benchmarks, including both classic and recently introduced datasets, researchers can assess whether improvements hold beyond data that may be contaminated. Ablation studies, meanwhile, demonstrate that performance gains stem from specific methods or frameworks rather than from data memorisation, since results drop when key components are removed.

Finally, reasoning-based safeguards, such as complex prompting~\cite{li2025coca} or zero-shot setups~\cite{zahan2025leveraging}, help reduce the likelihood that models succeed by recalling exact matches.

\faHandPointRight \ While no strategy can entirely eliminate the risk of contamination when the training dataset is unavailable, we recommend a combination of temporal filtering, benchmark design consideration, deduplication, cross-dataset validation, and ablation studies to greatly reduce its impact. These measures ensure results reflect model capabilities rather than memorisation and provide the community with practical standards for trustworthy LLM evaluation.

\noindent \textbf{RQ3: Replicability and Reproducibility}
\hfill \break
\indent \textit{\textbf{Model Configuration.}}
While the total number of studies reporting on inference (generation) configuration remains relatively low -- 89 out of 177 papers (50.3\%) -- we observe a positive trend over time: in 2023 10 out of 32 papers (31.2\%) included details about inference (generation) configuration; in 2024 this increased to 27 out of 55 papers (49.1\%), and by 2025 52 out of 90 papers (57.8\%). This upward trajectory suggests growing awareness of the importance of documenting inference settings, although almost half of the studies still omit them. Overall, we observe that model configurations are reported inconsistently across studies: some report only training parameters, some only generation parameters, and others report neither. While we acknowledge that page limits in conference venues make it difficult to document all details, the absence of such information significantly undermines replicability.

\faHandPointRight\ To mitigate these shortcomings, we recommend that empirical studies include, at a minimum, a permanent link to an external appendix or artefact repository where prompts (along with their rationale) and model configurations are described in detail. This practice would not only enhance transparency but also ensure that results can be meaningfully reproduced and built upon.

\textbf{\textit{Artefact Availability.}}
In our analysis, we find that 33 out of 177 (18.6\%) of empirical studies involving LLMs received an ``Artifact Available'' badge. This finding is in stark contrast to the remaining 515 articles published at ICSE, for which 213 badges were awarded (41.4\%), thus suggesting that authors conducting research involving LLMs have been less proactive in submitting their contributions to the artefact evaluation tracks. We observe similar trends for ``Artifact Reusable'' badges (11.9\% LLM-based vs. 29.1\% other) and ``Artifact Functional'' badges (7.3\% LLM-based vs. 13.6\% other). Over the three years examined, we observe a minor increase in the number of studies awarded availability badges. Per year, 6 of the 32 papers (18.75\%) from 2023 have these badges, 9 out of 55 papers (16.36\%) in 2024, and 18 out of 90 papers (20.00\%) in 2025.

We manually analysed the links to the artefacts provided in the examined papers and found that nine papers contain expired links. Of these nine, two had ``Artifact Available'' badges, and one even had an ``Artifact Reusable'' badge.

\faHandPointRight\ The community should actively promote the permanent availability of artefacts from LLM-based SE research and incentivise researchers to submit their work to artefact evaluation tracks. Such practices not only strengthen replicability and reproducibility but also foster transparency, enable fair comparisons, and accelerate cumulative progress in the field.

\noindent \textbf{RQ4: Sustainability}
\hfill \break
\indent \textbf{\textit{Reported Costs.}}
The most commonly reported cost across the 177 publications is the computational environment used to conduct experiments, as detailed in 89 papers (50.3\%). This entails the use of dedicated hardware, such as GPUs (77 times) or, in 2 cases, TPUs.

Time is the next most frequently discussed cost, being mentioned 68 times. This includes any reported notion of time (e.g., total execution time or the time required for a single LLM invocation). Thirty-six publications (20\%) reported both the computational environment and time costs, which are necessary to estimate the energy requirements for conducting experiments. This is an improvement over the 10.2\% (30 out of 293 publications) reported by Hort et al.~\cite{Hort2023}, which considered LLM studies on code-related tasks published up to 2022. However, we found no study that directly reported the energy consumed or $CO_2$ emissions.

Monetary costs have been reported 18 times, with budgets up to 3000\$. Twelve publications reported the sizes of input or output content (e.g., number of tokens or files) processed by LLMs, which dictate the incurred costs. Less frequent cost categories include memory requirements (6 times), number of LLM invocations (3 times), and number of floating point operations (1 time).

\faHandPointRight\ We encourage authors to clearly report the actual costs incurred to perform experiments, as well as estimates for the costs associated with using the proposed approach in practice over time. To raise awareness of sustainability matters, we believe this should become a standard practice required by conferences and journals, as is already the case for data availability.

\textbf{\textit{User Study.}}
We received 57 responses to our questionnaire. 95\% of participants used commercial models in the past 12 months, and 89\% said they are very likely (65\%) or likely (24\%) to use them in the next 12 months. However, the majority (56\%) do not know whether they will be able to sustain the costs over time, and 9\% believe that they will not be able to do so at all. In contrast, 82\% of participants used open models in the last 12 months, and 95\% said they are very likely (89\%) or likely (6\%) to use them in the next 12 months. Most participants (57\%) believe they will be able to sustain the hardware costs required to run open models over time, while 38\% are uncertain, and only one (2\%) says they will not be able to.

Participants reported spending as little as \$200 to as much as \$50,000 on commercial models in the last 12 months. For open-source models, researchers generally do not incur any direct costs. However, the hardware costs to run these models can be high, depending on usage goals (e.g., inference, training, fine-tuning), and our respondents reported spending from \$3,000 to \$40,000 to buy such hardware in the last year. Many mentioned electricity costs, yet none provided a monetary value, either because their institutions cover them or because they are difficult to measure.

When asked how they plan to sustain these costs in future, the participants gave very similar responses for commercial and open models: They plan to rely on a combination of academic funding such as research grants (36\% for commercial, 37\% for open), academic institutional support (26\% and 28\%), industrial support in the form of free credits and collaboration (24\% and 19\%), and some are willing to use their own personal funds (12\% and 9\%).

\faHandPointRight \ While researchers are adopting both commercial and open models, costs are a pressing concern. Nearly all participants used commercial models in the past year and intend to continue, yet most are unsure whether they can afford long-term usage due to API fees. Open models are also widely used, but researchers remain optimistic about sustaining their costs. Still, the bottleneck lies in the steep hardware and energy costs for inference and fine-tuning.

\section{Conclusions \& Future Plans}
The findings from our systematic review of ICSE publications highlight concerning trends in the reproducibility and sustainability of experiments using LLMs, echoing claims made in other meta-studies exploring the challenges of LLM-based research~\cite{sallou2024breaking, baltes2025guidelinesempiricalstudiessoftware}. 

In future work, we plan to deepen our analysis to cover additional empirical aspects, such as the clarity and rigour of reporting on prompts and datasets used. Further, we aim to tackle additional ethical and social factors, including whether the costs and infrastructure demands of LLMs are enabling a digital divide in the field, particularly between well-resourced and low-resourced institutions. 

Ultimately, we aim for our work to contribute to the emerging efforts by the wider community towards developing unified, actionable guidelines for the responsible and sustainable use of LLMs in software engineering research.

\begin{acks}
We thank the anonymous participants of our survey.

Max Hort is supported by the European Union through the Horizon Europe Marie Sk\l{}odowska-Curie Actions (\#101151798).

Aldeida Aleti is supported by the Australian Research Council, DP210100041.
\end{acks}

\newpage

\bibliographystyle{ACM-Reference-Format}
\bibliography{bibliography}

\end{document}